\documentclass[seceq,amsmath,subfigure,wrapfig,preprint]{ptptex}
\usepackage{graphicx}

\preprintnumber[3cm]{
OCU-PHYS-217\\AP-GR-17\\YITP-04-48\\PTPTeX ver.0.8\\ \today}

\title{
High-Speed Cylindrical Collapse of Perfect Fluid

}

\author{
Ken-ichi \textsc{Nakao}$^1$ and Yoshiyuki \textsc{Morisawa}$^2$
}

\inst{

Department of Physics,~Graduate School of Science,
~Osaka City University \\ Osaka 558-8585,~Japan\\
and\\
$^2$Yukawa Institute for Theoretical Physics, 
Kyoto University, Kyoto 606-8502, Japan}

\date{\today}

\abst{
The gravitational collapse of a cylindrically distributed 
perfect fluid is studied. We assume that the collapsing speed 
of the fluid is very large and investigate such 
a situation using recently proposed high-speed approximation scheme. 
We show that if the value of the pressure divided by the energy 
density is bounded below by some positive value, 
the high-speed collapse is necessarily  
decelerated by pressure repulsion so significantly that 
this approximation scheme becomes inapplicable. 
However, even in the case of a mono-atomic ideal gas, 
which is a typical example of the bounded case,
an arbitrarily large tidal force for freely falling observers 
can be realized before the high-speed approximation breaks down 
if the initial collapsing velocity is set to a very large value. 
By contrast, in case that the equation 
of state is sufficiently soft, we find that the high-speed collapse is 
maintained until a spacetime singularity forms. 
}
\begin{document}

\maketitle

\section{Introduction}

Gravitational collapse is an important topics in 
general relativity. With recent observational studies of gravitational 
waves using laser interferometric detectors such as LIGO, TAMA, VIRGO and 
GEO\cite{Ref:GW-US,Ref:GW-J}, the study of gravitational collapse 
as a source of gravitational radiation was gained its importance. 
Almost all theoretical studies on generation mechanisms of 
gravitational emissions have been proceeded under the 
assumption that the spacetime singularities are hidden 
inside black holes. However, if the spacetime singularities  
are not enclosed by event horizons, the situation could be 
completely different from the case with horizons. Such a spacetime 
singularity is called a globally naked 
singularity.\footnote{The visible spacetime singularity is called 
the naked singularity. The globally naked singularity is a special 
case of the naked singularity. It is defined as the naked singularity 
visible from infinity.}
Nakamura, Shibata and one of the present authors have 
conjectured that a large spacetime curvature in the neighborhood 
of a globally naked singularity can propagate away to infinity 
in the form of gravitational radiation due to the lack of an event
horizon and therefore the mass of the naked singularity is lost through 
large gravitational emissions\cite{Ref:DRIES-UP}.

Here we should note the cosmic censorship conjecture\cite{Ref:penrose69}. 
The statement of the cosmic censorship is, roughly speaking, 
that a spacetime singularity formed as the result of physically 
reasonable initial data cannot be naked. In connection with this matter, 
recently, Harada and one of the present authors proposed the 
alternative concept of the spacetime singularity, called a `spacetime 
border'\cite{Ref:Border}. There is a cut-off energy scale $\Lambda$ 
above which general relativity 
cannot describe any physical process; for example, 
$\Lambda$ could be the Planck scale $E_{\rm pl}$ ($\sim 10^{19}$GeV), 
or it might be a scale on the order of several TeV in the brane world 
scenario\cite{Ref:ADD,Ref:AADD,Ref:RS1,Ref:RS2}. 
Thus, spacetime regions of energy scales 
higher than $\Lambda$ should be regarded as singularities for  
general relativity. This viewpoint leads to the definition of 
a spacetime border as follows.  A spacetime border is defined as a 
region of the spacetime which satisfies the inequality  
\begin{equation}
\inf_{\cal A}F \ge {\Lambda^{4}\over E_{\rm pl}^{2}}~,
\label{eq:definition}
\end{equation}
where the curvature strength $F$ is given for instance by
\begin{equation}
F:=\max(|R^{a}{}_{a}|,|R^{ab}R_{ab}|^{1/2},|R^{abcd}R_{abcd}|^{1/2}, 
|R_{\mu\nu\rho\sigma}|).
\end{equation}
Here we have extended the original definition of $F$ to a slightly 
more general one. In this case, $F$ includes components 
$R_{\mu\nu\rho\sigma}$ of the Riemann tensor with respect to the 
tetrad basis transported parallelly along timelike 
curves of bounded acceleration, i.e., the tidal force experienced 
by physically reasonable observers. In a practical sense, 
the visible border is a useful concept as an alternative to 
the naked singularity. 

In the investigation of naked singularity formation, mainly 
spherically symmetric systems have been  
studied\cite{Ref:Joshi-text}, because they are 
simple but possess rich physical content. 
However, in a sense, such a system is too simple, as 
there is no degree of gravitational radiation. 
Therefore, in order to study the generation mechanism of gravitational 
radiation, we have to add non-spherical perturbations to 
this system\cite{Ref:IHN1,Ref:IHN2,Ref:IHN3,Ref:NIH,Ref:HIN} 
or consider non-spherically symmetric spacetimes. 
In this sense, cylindrically symmetric gravitational 
collapse has significant physical meaning, because 
there is a degree of gravitational radiation, and, further, 
the spacetime singularity in this system is 
naked\cite{Ref:HOOP,Ref:Hayward}. 
There are a few numerical studies of gravitational 
wave generation through cylindrical gravitational 
collapse\cite{Ref:Piran,Ref:Dust-Shell,Ref:Chiba}. Although there 
has been debate about their results, the present authors have 
revealed with an analytic study that there is no 
inconsistency\cite{Ref:paperI}. 
In any case, these investigations give strong evidence for the 
generation of a large amount of gravitational radiation through  
cylindrical gravitational collapse. There are also several 
studies concerning cylindrical gravitational collapse, 
and these have improved our understanding of non-spherical relativistic 
dynamics\cite{Ref:Morgan,Ref:AT,Ref:LW,Ref:Nolan,Ref:PW,Ref:N&N}. 

In this paper, we investigate the gravitational collapse 
of a cylindrical, thick shell composed of a perfect fluid 
by generalizing a high-speed approximation for a cylindrical 
thick dust shell\cite{Ref:paperI} to the case of non-vanishing pressure. 
In this approximation scheme, the collapsing speed 
is assumed to be almost equal to the speed of light. Therefore we 
treat the deviation of the 4-velocity of the perfect fluid from 
null as a small perturbation. By using this approximation scheme, 
we study the effect of the pressure in the case of a very  
large collapsing velocity. 

This paper is organized as follows. In $\S$2, we briefly review the 
formulation of a spacetime with whole-cylinder 
symmetry\cite{Ref:Melvin}. Then in $\S$3, we present 
the high-speed approximation scheme for a general perfect fluid. In $\S$4, 
we discuss the effect of pressure and determine whether the pressure 
prevents high-speed collapse. Because an ideal mono-atomic 
gas behaves like radiation in the high energy limit, in $\S$5 
we solve the Euler equation with a linear equation of state 
including the radiation case, and show that if 
the initial collapsing velocity is chosen appropriately, 
then an arbitrarily strong tidal force for freely falling observers 
is realized in the neighborhood of the symmetric axis  
of the cylinder before the high-speed approximation 
becomes inapplicable. Finally, $\S$6 is devoted to 
a summary and discussion. 

In this paper, we adopt $c=1$ unit and basically follow the convention 
and notation in the textbook by Wald\cite{Ref:Wald}.

\section{Cylindrically symmetric perfect fluid system}

We focus on a spacetime of whole-cylinder symmetry. Such a system
possesses the line element\cite{Ref:C-energy}
\begin{equation}
ds^{2}=e^{2(\gamma-\psi)}\left(-dt^{2}+dr^{2}\right)
+e^{2\psi}dz^{2}+e^{-2\psi}R^{2}d\varphi^{2},
\end{equation}
where the metric variables $\gamma$, $\psi$ and $R$ are functions of 
$t$ and $r$. Then, the Einstein equations are
\begin{eqnarray}
&&\gamma'=\left({R'}^{2}-{\dot R}^{2}\right)^{-1}
\biggl\{
RR'\left({\dot \psi}^{2}+{\psi'}^{2}\right)
-2R{\dot R}{\dot \psi}\psi'
+R'R''-{\dot R}{\dot R}' \nonumber \\ 
&&~~~~-8\pi G\sqrt{-g}\left(R'T^{t}{}_{t}+{\dot R}T^{r}{}_{t}\right)
\biggr\}, 
\label{eq:einstein-1} \\
&&{\dot \gamma}=-\left({R'}^{2}-{\dot R}^{2}\right)^{-1}
\biggl\{
R{\dot R}\left({\dot \psi}^{2}+{\psi'}^{2}\right)
-2RR'{\dot \psi}\psi' 
+{\dot R}R''-R'{\dot R}' \nonumber \\
&&~~~~-8\pi G\sqrt{-g}\left({\dot R}T^{t}{}_{t}+R'T^{r}{}_{t}\right)
\biggr\}, \\
&&{\ddot \gamma}-\gamma''={\psi'}^{2}-{\dot \psi}^{2}-{8\pi G\over R}
\sqrt{-g}T^{\varphi}{}_{\varphi}, \\
&&{\ddot R}-R''
=-8\pi G\sqrt{-g}\left(T^{t}{}_{t}+T^{r}{}_{r}\right), \\
&&{\ddot \psi}+{{\dot R}\over R}{\dot \psi}-\psi''
-{R'\over R}\psi' 
=-{4\pi G\over R}\sqrt{-g}\left(T^{t}{}_{t}+T^{r}{}_{r}-T^{z}{}_{z}
+T^{\varphi}{}_{\varphi}\right), 
\label{eq:einstein-2}
\end{eqnarray}
where the dot represents differentiation with respect to $t$ and the 
prime that with respect to $r$. 

In this paper, we consider a perfect fluid. The stress-energy tensor 
is written 
\begin{equation}
T_{\mu\nu}=(\rho+p)u_{\mu}u_{\nu}+pg_{\mu\nu},
\end{equation}
where $g_{\mu\nu}$ is the metric tensor, $\rho$ is the energy 
density, $p$ is the pressure, and $u^{\mu}$ is the 4-velocity of 
the fluid element.  Due to the assumption of the whole-cylinder 
symmetry, the components of the 4-velocity $u^{\mu}$ are written 
\begin{equation}
u^{\mu}=u^{t}\left(1,-1+V,~0,~0\right),
\end{equation}
where $V$ should be positive so that $u^\mu$ is timelike. 
With the normalization $u^{\mu}u_{\mu}=-1$, $u^{t}$ is expressed as 
\begin{equation}
u^{t}={e^{-\gamma+\psi}\over \sqrt{V\left(2-V\right)}}.
\end{equation}

Here we introduce the new variables $D$ and $P$ defined by
\begin{eqnarray}
D&:=&{\sqrt{-g}(\rho+p) u^{t}\over \sqrt{V\left(2-V\right)}}
={R e^{\gamma-\psi}(\rho+p) \over V\left(2-V\right)}, 
\label{eq:D-def}\\
P&:=&{R e^{\gamma-\psi}p \over V\left(2-V\right)}, 
\label{eq:P-def}
\end{eqnarray}
where $g$ is the determinant of the metric tensor $g_{\mu\nu}$. 
The components of the stress-energy tensor $T^\alpha{}_\beta$ 
are then expressed as
\begin{eqnarray}
\sqrt{-g}T^{t}{}_{t}&=&e^{\gamma-\psi}\left\{-D+V\left(2-V\right)P\right\}, \\
\sqrt{-g}T^{r}{}_{t}&=&e^{\gamma-\psi}D(1-V)=-\sqrt{-g}T^{t}{}_{r}, \\
\sqrt{-g}T^{r}{}_{r}&=&e^{\gamma-\psi}\left\{\left(1-V\right)^{2}D
                     +V\left(2-V\right)P\right\}, \\
\sqrt{-g}T^{z}{}_{z}&=&\sqrt{-g}T^{\varphi}{}_{\varphi}=e^{\gamma-\psi}V(2-V)P,
\end{eqnarray}
and the other components vanish. 

The equation of motion $\nabla_{\alpha}T^{\alpha}{}_{\beta}=0$ leads to 
\begin{eqnarray}
\partial_{u}D&=&-{1\over2}(DV)'+{1\over 2}\left\{V(2-V)P\right\}{\dot{} }
+{D\over2}(1-V)\left\{2\partial_{u}(\psi-\gamma)
-V({\dot \psi}-{\dot\gamma})\right\} \nonumber \\
&+&{1\over2}PV(2-V)({\dot \psi}-{\dot\gamma}-{\dot R}/R), 
\label{eq:conservation-full}\\
&& \nonumber \\
D\partial_{u}V&=&(1-V)\partial_{u}D+{1\over2}\left\{V(1-V)D-V(2-V)P\right\}'
\nonumber\\
&-&{D\over2}\left\{2\partial_{u}(\psi-\gamma)-V({\dot \psi}-{\dot\gamma})\right\} 
+{1\over2}PV(2-V)(\gamma'-\psi'+R'/R),  \label{eq:Euler-full}
\end{eqnarray}
where $u=t-r$ is the retarded time and $\partial_{u}$ is the partial 
derivative of $u$ with fixed advanced time $v=t+r$. The first equation
comes from the $t$-component, while the second one comes from the
$r$-component. The $z$- and $\varphi$-components are trivial.  

To estimate the energy flux of gravitational radiation, 
we adopt the $C$-energy $E$ and its flux vector proposed by 
Thorne\cite{Ref:C-energy}. The $C$-energy $E=E(t,r)$ is the 
energy per unit coordinate length along the $z$-direction 
within a radius $r$ at time $t$. It is defined by 
\begin{equation}
E={1\over8}\left\{1+e^{-2\gamma}\left({\dot R}^2-{R'}^2\right)\right\}.
\end{equation}
The energy flux vector $J^\mu$ associated with $C$-energy is defined by
\begin{equation}
\sqrt{-g}J^\mu=\left(
{\partial E\over \partial r},-{\partial E\over \partial t},
0,0\right).
\end{equation}
By definition, $J^{\mu}$ is divergence free. Using the equations of 
motion for the metric variables, we obtain the expression for 
the $C$-energy flux vector as 
\begin{eqnarray}
\sqrt{-g}J^{t}&=&{e^{-2\gamma}\over 8\pi G}
\biggl\{RR'({\dot \psi}^{2}+{\psi'}^{2}) 
-2R{\dot R}{\dot \psi}\psi'
-8\pi G\sqrt{-g}(R'T^{t}{}_{t}+{\dot R}T^{r}{}_{t})\biggr\}, \\
\sqrt{-g}J^{r}
&=&{e^{-2\gamma}\over 8\pi G}
\biggl\{R{\dot R}({\dot \psi}^{2}+{\psi'}^{2}) 
-2RR'{\dot \psi}\psi' 
-8\pi G\sqrt{-g}(R'T^{r}{}_{t}-{\dot R}T^{r}{}_{r})\biggr\},
\end{eqnarray}
while the other components vanish.

\section{High-speed approximation for a perfect fluid} 

Let us consider the ingoing null limit of the cylindrical perfect 
fluid. Using $D$ and $P$, the stress-energy tensor is written  
\begin{equation}
T_{\mu\nu}={e^{3(\psi-\gamma)}\over R}\left\{D k_{\mu}k_{\nu}
+e^{2(\gamma-\psi)}V(2-V)Pg_{\mu\nu}\right\},
\end{equation}
where
\begin{equation}
k^{\mu}=(1,-1+V,~0,~0).
\end{equation}
The timelike vector $k^{\mu}$ becomes an ingoing null vector 
in the limit $V\rightarrow0_+$.  
Hence, in this limit with $D$ and $P$ fixed, 
the stress-energy tensor is identical to that of collapsing null dust, 
\begin{equation}
T_{\mu\nu} \longrightarrow {e^{3(\psi-\gamma)}D \over R}k_{\mu}k_{\nu},
\label{eq:n-limit-st-tensor}
\end{equation}
where
\begin{equation}
k^{\mu} \longrightarrow (1,-1,~0,~0).
\end{equation}
This implies that in the case of a very large collapsing velocity, 
i.e., for $0<V\ll1$, the perfect fluid system is 
approximated well by a null dust system. For this reason, we treat the 
deviation $V$ of the 4-velocity from null 
as a perturbation and perform a linear perturbation analyses. 

In the case of collapsing null dust, the solution is 
easily obtained as 
\begin{eqnarray}
\psi&=&0, \\
\gamma&=&\gamma_{B}(v), \label{eq:gamma-BG}\\
R&=&r, \label{eq:R-BG} \\
8\pi GDe^{\gamma}&=&{d\gamma_{B}\over dv}, \label{eq:D-BG}
\end{eqnarray}
where $\gamma_{B}(v)$ is an arbitrary function of the 
advanced time $v$. 
This solution was obtained by Morgan\cite{Ref:Morgan} and was studied 
subsequently by Letelier and Wang\cite{Ref:LW} and Nolan\cite{Ref:Nolan} 
in detail. We regard this solution as a background spacetime for 
the perturbation analysis. The situation can be understood from Fig.1. 
The ``density'' $D$ is assumed to have compact support, 
$0<v<v_{\rm w}$, which is depicted by the shaded region. 
We find from Eqs.(\ref{eq:n-limit-st-tensor}) and 
(\ref{eq:R-BG}) that if $D$ does not vanish at the symmetric axis 
$r=0$, the components of the stress-energy tensor $T_{\mu\nu}$ with respect 
to the coordinate basis diverge there, and the same is true for 
the Ricci tensor by Einstein equations. 
This is a naked singularity which is depicted 
by the dashed line at $r=0$ on the interval $0< t< v_{\rm w}$ in Fig.1. 
Although all the scalar polynomials of the Riemann tensor vanish there, 
freely falling observers experience an infinite tidal force at $r=0$ for 
$0<t<v_{\rm w}$\cite{Ref:LW}. The short dashed line corresponding to 
$t=r$ represents 
the Cauchy horizon associated with this naked singularity. 
The region satisfying 
$t\geq v_{\rm w}$ at $r=0$ is a conical singularity 
which is depicted by the dotted line in Fig.1. 

\begin{figure}
\rotatebox{0}{\resizebox{10cm}{!}{\includegraphics{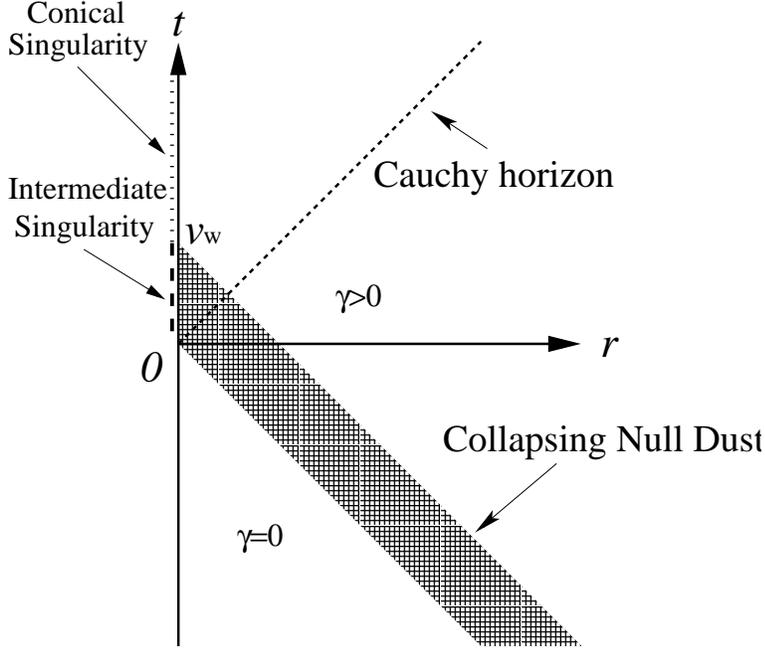}}}
\caption{
Morgan's cylindrical null dust solution. 
Null dust exists in the shaded region. In this situation depicted, 
$D(v)>0$ for $0<v<v_{\rm w}$. The dashed line on the $t$-axis corresponds to 
the intermediate singularity at which an observer experiences an infinite 
tidal force, although the scalar polynomials of the Riemann tensor 
do not vanish.  The dotted line on $t$-axis represents 
the conical singularity. The Cauchy horizon is represented by the 
short dashed line corresponding to $t=r$. 
}
\label{fg:morgan}
\end{figure}

Now let us consider a general perfect fluid with very large collapsing 
velocity. To carry out the linear perturbation 
analysis, we introduce a small parameter $\epsilon$ and assume 
the orders $V= O(\epsilon)$ and $\psi=O(\epsilon)$.
Further, we rewrite the variables $\gamma$, $R$ and $D$ as
\begin{eqnarray}
e^{\gamma}&=&e^{\gamma_{B}}(1+\delta_{\gamma}), \\
R&=&r(1+\delta_{R}), \\
D&=&D_{B}(1+\delta_{D}),
\end{eqnarray}
and assume that $\delta_{\gamma}$, $\delta_{R}$ 
and $\delta_{D}$ are $O(\epsilon)$, where
\begin{equation}
D_{B}:={1\over 8\pi Ge^{\gamma_{B}}}{d\gamma_{B}\over dv}.
\label{eq:DB-def}
\end{equation}
We call the perturbative analysis with respect to this small
parameter $\epsilon$ the `high-speed approximation'. 

The first-order equations with respect to $\epsilon$ are 
given as follows: the Einstein equations 
(\ref{eq:einstein-1})$-$(\ref{eq:einstein-2}) lead to
\begin{eqnarray}
&&\delta_{\gamma}{}'=
8\pi GD_{B}e^{\gamma_{B}}\left\{\delta_{\gamma}-\psi+\delta_{D} 
-2\partial_{v}(r\delta_{R})-2PV/ D_{B}\right\} 
+(r\delta_{R})'',
\label{eq:del-r-gamma}\\
&&{\dot \delta}_{\gamma}=
8\pi GD_{B}e^{\gamma_{B}}\left\{\delta_{\gamma}-\psi+\delta_{D}
-2\partial_{v}(r\delta_{R})-V\right\} +(r{\dot\delta_{R}})', 
\label{eq:del-t-gamma}\\
&&{\ddot\delta}_{\gamma}-\delta_{\gamma}{}''
=-{16\pi G \over r}e^{\gamma_{B}}PV, 
\label{eq:gamma-linear}\\
&&r{\ddot\delta}_{R}-(r\delta_{R})''=16\pi Ge^{\gamma_{B}}(D_{B}-2P)V, 
\label{eq:R-linear}\\
&&{\ddot\psi}-\psi''-{1\over r}\psi'={8\pi G \over r}
e^{\gamma_{B}}(D_{B}-2P)V; 
\label{eq:psi-linear}
\end{eqnarray}
the conservation law (\ref{eq:conservation-full}) leads to  
\begin{eqnarray}
\partial_{u}\left(\delta_{D}+\delta_{\gamma}-\psi\right)
&=&-{1\over 2D_{B}}{dD_{B}\over dv}V 
-{1\over2}\left(V'-{d\gamma_{B}\over dv}V\right) 
\nonumber \\
&&~~~~~~~~~~~~~~~~~~
+{1\over D_{B}}\left\{(PV){\dot{}}
-PV{d\gamma_{B}\over dv}\right\};
\label{eq:conservation-linear}
\end{eqnarray}
the Euler equation (\ref{eq:Euler-full}) becomes 
\begin{equation}
\partial_{u}\left\{(D_{B}-2P)V\right\}={PV\over r},  \label{eq:V-linear}
\end{equation}
where we have used Eq.(\ref{eq:conservation-linear}). 

The $C$-energy $E$ up to first order in $\epsilon$ is given by
\begin{equation}
E={1\over8}\left[1-e^{-2\gamma_B}+2e^{-2\gamma_B}
\left\{\delta_\gamma-\left(r\delta_R\right)'\right\}\right].
\label{eq:C-energy-linear}
\end{equation}
We can easily see in Eq.(\ref{eq:DB-def}) that $\gamma_B$ 
is constant in the vacuum region, $D_B=0$. 
Further, from Eqs.(\ref{eq:del-r-gamma}) and (\ref{eq:del-t-gamma}), 
we find that $\delta_\gamma-(r\delta_R)'$ is constant in the vacuum region. 
Therefore, up to first order, the $C$-energy is constant 
in the vacuum region. This means that in the vacuum region, 
the $C$-energy flux vector $J^\mu$ vanishes up to this order, 
and thus it is a second-order quantity given by
\begin{eqnarray}
\sqrt{-g}J^{t}&=&{r\over 8\pi G}
\left({\dot \psi}^{2}+{\psi'}^{2}\right), \\
\sqrt{-g}J^{r}&=&-{r\over 4\pi G}{\dot \psi}\psi'. 
\label{eq:C-flux-linear}
\end{eqnarray}
In this case, the $C$-energy flux vector takes a form very similar 
to that of the massless Klein-Gordon field. 

\section{Does pressure prevents the high-speed collapse?}

Assuming that both the energy density $\rho$ and pressure $p$ of the 
fluid are positive, we introduce the quantity 
\begin{equation}
c_{\rm s}(u,v)=\sqrt{p\over\rho}  \label{eq:cs-def}
\end{equation}
If the dominant energy condition is satisfied, 
$c_{\rm s}$ is smaller than or equal to the speed of light\cite{Ref:Wald}. 
However, in this paper, we assume a more stringent constraint 
on $c_{\rm s}$, namely 
\begin{equation}
c_{\rm s}<1.
\end{equation}
As shown below, the present approximation scheme 
cannot treat the case of $c_{\rm s}=1$. 

From Eqs.(\ref{eq:D-def}) and (\ref{eq:P-def}) and by using 
Eq.(\ref{eq:cs-def}), we obtain 
\begin{equation}
P={c_{\rm s}{}^{2}\over 1+c_{\rm s}{}^{2}}D. \label{eq:EOS}
\end{equation}
With the above equation, the Euler equation (\ref{eq:V-linear}) 
can be rewritten as 
\begin{equation}
\partial_u\ln\left({1-c_{\rm s}{}^2\over 1+c_{\rm s}{}^2}V\right)
={c_{\rm s}{}^2\over (1-c_{\rm s}{}^2)r}.
\label{eq:V-linear-2}
\end{equation}
Integrating this equation formally, we obtain 
\begin{equation}
V={1+c_{\rm s}{}^2\over 1-c_{\rm s}{}^2}
\exp\left[\int^u_{U(v)}dx{2c_{\rm s}{}^2(x,v)
\over\{1-c_{\rm s}{}^2(x,v)\}(v-x)}\right],
\label{eq:V-formal}
\end{equation}
where $U(v)$ is an arbitrary function of the advanced time $v$. 
In the above equation, the velocity perturbation $V$ is 
not defined for $c_{\rm s}=1$, and thus the high-speed 
approximation is not applicable to this case. 

We consider two cases, one in which 
$c_{\rm s}$ is bounded below by some positive constant, 
and one in which $c_{\rm s}$ vanishes in 
the large energy density limit, $\rho\rightarrow\infty$.  

\subsection{Bounded case}

In case that $c_{\rm s}$ is bounded below by some positive value, 
we have the inequality   
\begin{equation}
{2c_{\rm s}{}^2\over 1-c_{\rm s}{}^2}\geq b^2,
\end{equation}
where $b$ is some non-vanishing constant.  
Then from Eq.(\ref{eq:V-linear-2}), we obtain
\begin{equation}
\partial_u\ln\left({1-c_{\rm s}{}^2\over 1+c_{\rm s}{}^2}V\right)
\geq{b^2\over v-u}.
\end{equation}
Integrating the above inequality, we obtain
\begin{equation}
V\geq {\rm Const}\times r^{-b^2},
\end{equation}
where `Const' should be positive. 
We find in the above inequality that the velocity perturbation 
$V$ diverges when the fluid element approaches the 
symmetric axis, $r=0$. This implies that in this case, 
the high-speed collapse is necessarily decelerated so significantly that 
the high-speed approximation breaks down.  

\subsection{Vanishing case}

If $c_{\rm s}$ vanishes in the limit of a high energy 
density, $\rho \rightarrow \infty$, the behavior of the perfect fluid 
can differ from the case of bounded $c_{\rm s}$. 
When the fluid elements approach the background singularity, 
i.e., $r=0$ in the region of non-vanishing $D_B$, 
the energy density will become larger and larger.  
Thus, in this case, the asymptotic behavior of $c_{\rm s}$ 
near the background singularity will be given by 
\begin{equation}
c_{\rm s}{}^2(u,v)\sim C_{\rm s}{}^2(t)r^q 
=C_{\rm s}{}^2(v-r)r^q\sim C_{\rm s}{}^2(v)r^q, 
\label{eq:cs-asymp}
\end{equation}
where $C_{\rm s}$ is some function, and $q$ is a positive constant. 
Substituting the above 
asymptotic form of $c_{\rm s}$ into Eq.(\ref{eq:V-formal}), we obtain
\begin{equation}
V\sim C_V{}^2(v)
\exp\left\{{2^{q+1}\over q}C_{\rm s}{}^2(v)r^q\right\}, 
\label{eq:V-asymptotic-1}
\end{equation}
where
\begin{equation}
C_V(v)=\exp\left[-{C_{\rm s}{}^2(v)\over q}\left\{v-U(v)\right\}^q\right].
\end{equation}
Note that $C_V$ should have the same support as the background 
density variable $D_B$. Therefore, in this case, the velocity 
perturbation remains finite, even at the background singularity. 
This seems to imply that the naked singularity forms at the background 
singularity through this gravitational collapse. However,  
the full-order analysis is necessary to reach 
a definite conclusion.

Here, conversely, we consider the asymptotic equation of state that 
realizes the asymptotic behavior (\ref{eq:cs-asymp}). We find from 
Eq.(\ref{eq:D-def}) that as the background singularity is approached, 
the energy density behaves as 
\begin{equation}
\rho \sim {2VD_B\over re^{\gamma_B}} 
=({\rm function~~of}~~v)^2 \times {1\over r}, 
\end{equation}
where we have used the fact that $V$ becomes a function of the advanced 
time $v$ at $r=0$, as seen from Eq.(\ref{eq:V-asymptotic-1}). 
Therefore together with Eqs.(\ref{eq:cs-def}) and (\ref{eq:cs-asymp}), 
the above equation leads to 
\begin{equation}
p=c_{\rm s}{}^2\rho\sim {\rm Const}\times \rho^{1-q}
\end{equation}
near $r=0$, where `Const' should be positive. This result suggests that 
a very soft equation of state is necessary so that 
the high-speed collapse is not halted by the pressure 
before spacetime singularity formation. 

\subsection{Generation of gravitational waves}

Equation (\ref{eq:psi-linear}) implies that a 
larger velocity perturbation $V$ leads to a larger $|\psi|$. 
Because the $C$-energy flux in the asymptotically flat region is given by 
Eq.(\ref{eq:C-flux-linear}), this suggests a 
large gravitational emission by large $V$. 

In the bounded $c_{\rm s}$ case, the growth of the velocity perturbation 
$V$ is unbounded as the  background singularity is approached. Although a 
velocity perturbation $V$ larger than unity is meaningless in the 
present approximation scheme, this result suggests that 
a significant amount of gravitational radiation is generated 
by pressure deceleration near the background singularity.  
This is consistent with Piran's numerical 
results, because Piran's numerical simulations for a cylindrically 
distributed ideal gas show that a large amount of gravitational 
radiation is emitted when the pressure bounce occurs. 
However, here we stress that in this paper, we study the result 
for an arbitrarily large initial collapsing velocity of the fluid, 
while Piran's numerical study is restricted to cases with 
momentarily static initial data. 

In the case of vanishing $c_{\rm s}$, the velocity perturbation takes 
a finite value even at the background singularity. Therefore the 
situation here is similar to that for a dust fluid in the 
generation processes of gravitational waves\cite{Ref:paperI}. 

\section{Appearance of the spacetime border}

When the gravitational collapse realizes a very high energy 
density, the fluid will behave as a mono-atomic ideal gas, due to 
the asymptotic freedom of the elementary interactions.  
Furthermore, the equation of state will 
be similar to that of the radiation, i.e., with $p=\rho/3$. 
This corresponds to the bounded $c_{\rm s}$ case, 
and the high-speed collapse is necessarily 
decelerated so significantly that the high-speed approximation 
becomes inapplicable before the fluid elements reach the background singularity. 
This result seems to imply that the formation of 
spacetime singularities is prevented by the occurrence of 
the pressure bounce. However, even if this is the case, as mentioned in $\S$1, 
it is physically important in a practical sense only whether 
the spacetime border forms before the occurrence of the pressure bounce. 
Therefore, here we consider the formation of the spacetime border in the bounded 
$c_{\rm s}$ case. 

We consider a linear equation of state, i.e., 
$c_{\rm s}=$constant, which includes the case of 
radiation, $c_{\rm s}{}^2=1/3$. Using Eq.(\ref{eq:EOS}), 
the Euler equation (\ref{eq:V-linear}) becomes
\begin{equation}
\partial_{u}V
={2c_{\rm s}{}^{2}V \over \left(1-c_{\rm s}{}^{2}\right)(v-u)}.
\label{eq:isothermal-V-linear}
\end{equation}
As mentioned previously, the above equation is singular for 
stiff matter with $c_{\rm s}=1$. The case $c_{\rm s}=1$ 
corresponds to a massless Klein-Gordon field, 
and it is easy to see that no gravitational radiation is 
generated in the cylindrically symmetric case. 

Equation (\ref{eq:isothermal-V-linear}) can be easily integrated, 
and we obtain
\begin{equation}
V=C_L(v)r^{-2c_{\rm s}{}^{2}/(1-c_{\rm s}{}^{2})},
\label{eq:V-isothermal}
\end{equation}
where $C_L(v)$ is an arbitrary positive function of 
the advanced time $v$, and it should have the same support 
as the background density variable $D_B$. As discussed in the 
preceding section, 
$V$ diverges to $+\infty$ at the background singularity 
in the case that $0<c_{\rm s}<1$. 
This implies that the perturbation analysis necessarily breaks down 
in the neighborhood of the background singularity. 

Once we know the velocity perturbation $V$, we can easily 
obtain solutions for $\delta_\gamma$, $\delta_R$ and $\psi$  
from Eqs.(\ref{eq:del-r-gamma})$-$(\ref{eq:psi-linear}) using the 
ordinary procedure to construct solutions of wave equations 
in flat spacetimes. 
We can also easily obtain a solution for $\delta_D$ by solving 
Eq.(\ref{eq:conservation-linear}). However, in this paper, 
we do not construct the solutions for these variables 
but instead focus on the velocity perturbation $V$ only. 

The high-speed approximation scheme 
is applicable only to situations in which $V\ll1$. 
Therefore, Eq.(\ref{eq:V-isothermal}) leads to a 
necessary condition for the applicability of  
high-speed approximation,
\begin{equation}
r \gg C_L{}^{(1-c_{\rm s}{}^2)/2c_{\rm s}{}^2}, 
\end{equation}
where $r$ is the radial position of a fluid element. 
It should be noted that if the arbitrary function $C_L$ is chosen  
sufficiently small, the present approximation scheme 
may be applicable even in the case of very small $r$. 
Because the tidal force experienced by freely 
falling observers diverges as 
the background singularity at $r=0$ is approached\cite{Ref:LW}, 
the velocity perturbation $V$ remains sufficiently small 
for the case of an arbitrarily large tidal force 
for freely falling observers if $C_L$ is chosen sufficiently small.  
This implies that if the initial collapsing velocity is sufficiently large, 
the gravitational collapse of a cylindrically distributed 
perfect fluid with a linear equation of state forms a spacetime border. 
Because there is no event horizon in the present case, 
the spacetime border is necessarily visible.   
This result implies practically the violation of cosmic 
censorship.  

\section{Summary and discussion}

In this paper, we have presented a high-speed approximation scheme 
applicable to a cylindrically symmetric perfect fluid. 
Within this approximation scheme, we have found that 
in the case of $c_{\rm s}$ bounded  below by some positive value, 
the high-speed collapse of the perfect fluid is necessarily 
decelerated so significantly that the 
high-speed approximation becomes inapplicable. 
This result reveals the possibility of pressure bounces, 
because the deceleration of the 
collapsing velocity implies that the pressure repulsion 
overwhelms the gravitational attraction. 
If this expectation holds, then the spacetime singularity formation 
resulting from the adiabatic collapse of a cylindrical perfect fluid 
of a physically reasonable ideal gas is impossible. 
However, here we should note that if the initial collapsing velocity 
is very large, a spacetime border (i.e., a region with 
spacetime curvature so large that the quantum effects in gravity 
are important) is realized before the high-speed approximation scheme 
breaks down. Because a horizon does not form in the system 
considered here, the spacetime border is necessarily visible. 
Therefore, this result implies practically the violation 
of cosmic censorship. By contrast, in the case that 
$c_{\rm s}$ vanishes in the limit of high energy density, 
$\rho\rightarrow\infty$, 
the high-speed collapse continues until a naked singularity forms. 
Thus, it is likely that if the equation of state is sufficiently soft, 
so that $p$ is proportional to $\rho^{1-q}$ with $0<q<1$, 
high-speed collapse is not prevented by the effect of the pressure, 
and thus a naked singularity forms. 

Equation (\ref{eq:psi-linear}) implies that a large velocity 
perturbation $V$ leads to the efficient generation of gravitational 
radiation. 
Therefore, the unlimited growth of $V$ in the bounded $c_{\rm s}$ case   
implies the generation of significant gravitational radiation. 
Contrastingly, in the case that $c_{\rm s}$ vanishes  
in the high energy density limit, $V$  
is finite, even at the background singularity. 
In this case, the behavior of the perfect fluid will be similar to 
that of a dust fluid, in which case a thinner cylindrical shell 
generates the larger amount of gravitational radiation\cite{Ref:paperI}. 

In the case that $c_{\rm s}$ vanishes in the high energy 
density limit, the gravitational emission from the 
naked singularity might have some physical importance. 
In the present approximation 
scheme, the metric variables are regarded as test fields in 
the background spacetime. Thus, if we appropriately impose 
boundary conditions on these variables at the background singularity, 
we can obtain solutions of the linearized Einstein equations and estimate 
the energy carried away by the gravitational waves from the naked 
singularity in this case. This topics will be discussed 
elsewhere.\cite{Ref:paperIII}

The approximation scheme used here is a kind of perturbation method. 
Strictly speaking, the full-order analysis is necessary 
to reach a definite conclusion, although 
the results up to first order might have significant physical
meaning. Therefore, a higher-order analysis should be meaningful, 
but this is a future work.

\section*{Acknowledgements}
We are grateful to our colleagues in the astrophysics and 
gravity group of Osaka City University for helpful discussions and criticism.  
KN thanks A. Hosoya and A. Ishibashi for useful discussion. 
This work is supported by a Grant-in-Aid for Scientific Research 
(No.16540264) from JSPS and also by a Grant-in-Aid for the 21st Century 
COE ``Center for Diversity and Universality in Physics'', 
from the Ministry of Education, Culture, Sports, Science and 
Technology (MEXT) of Japan.

%

\end{document}